\documentstyle[12pt,fleqn]{article}
\def\ref{par\noindent\hangindent=6mm\hangafter=1}
\begin{document}
\baselineskip 8mm

\def\f{factorization }
\def\p{polynomial }
\def\pp{polynomials }

\begin{center}
{\bf Recurrence-shift relations for the polynomial functions of Aldaya,
Bisquert, and Navarro-Salas}

\bigskip

H.C. Rosu\footnote{Electronic-mail: rosu@ifug.ugto.mx}$^{\dagger\pounds}$,
M.A. Reyes\footnote{Electronic-mail: mareyes@fnalv.fnal.gov}
$^{\ddagger}$
and O. Obreg\'on\footnote{Electronic-mail: octavio@ifug.ugto.mx}
$^{\dagger\P}$


$^{\dagger}$
Instituto de F\'{\i}sica de la Universidad de Guanajuato,
Apdo Postal E-143, Le\'on, Gto., M\'exico

$^{\pounds}$
Institute of Gravitation and Space Sciences, P.O. Box MG-6, Magurele-Bucharest,
Romania


$^{\ddagger}$
Departamento de F\'{\i}sica, Centro de Investigaci\'on y Estudios
Avanzados del Instituto Polit\'ecnico Nacional, Apdo Postal 14-740,
M\'exico Distrito Federal, M\'exico


$^{\P}$
Universidad Aut\'onoma Metropolitana, Iztapalapa, Apdo Postal 55-534,

M\'exico Distrito Federal, M\'exico

\end{center}
\bigskip
\bigskip

\newpage
\begin{abstract}
Using a simple factorization scheme we obtain the recurrence-shift
relations of the
polynomial functions of Aldaya, Bisquert and
Navarro-Salas (ABNS), $F_n^N(\frac{\omega}{c}\sqrt{N}x)$, i.e., one-step
first-order differential relations
referring to $N$, as follows. Firstly, we apply the scheme to the \p degree
confirming the recurrence relations of Aldaya, Bisquert and Navarro-Salas,
but also obtaining another slightly modified pair. Secondly,
the factorization scheme is applied to the Gegenbauer
polynomials to get the recurrence relations with respect to their parameter.
Next, we make use of Nagel's result, showing the connection between
Gegenbauer polynomials and the ABNS functions,
to write down the recurrence-shift relations for the latter ones.
Such relations may be used in the study of the spatial structure of
pair-creation processes in an Anti-de Sitter gravitational background.

\end{abstract}
\bigskip
\begin{center}
{\bf Resumen}

Usando un esquema de factorizaci\'on simple, obtenemos relaciones de
recurrencia de los polinomios usados por Aldaya,
Bisquert y Navarro-Salas para describir osciladores harmonicos relativistas,
$F_n^N(\frac{\omega}{c}\sqrt{N}x)$, incluyendo un
desplazamiento discreto en el argumento, es decir,
relaciones diferenciales de primer orden
en un paso con respecto a $N$, como sigue. Primeramente aplicamos el esquema al
grado del polinomio confirmando las relaciones de recurrencia de Aldaya et al.,
pero tambien, obtenemos otro par ligeramente modificado; ademas para los
polinomios de Gegenbauer obtenemos  una relaci\'on de recurrencia con respecto
a sus par\'ametro. Enseguida, haciendo uso de la relaci\'on de Nagel entre
los polinomios de Gegenbauer y las funciones ABNS obtenemos un par de
relaci\'ones de recurrencia-desplazamiento para esas funciones.
Tales relaciones podrian ser usados en el estudio de la
estructura espacial de procesos de creaci\'on de pares en un fondo
gravitacional de Anti-de Sitter.
\end{center}

\newpage



\section*{1.~Introduction}

Some time ago, Aldaya, Bisquert and Navarro-Salas (ABNS) \cite{1}
discussed a relativistic
generalization of the quantum harmonic oscillator for which the spacing of
the energy levels is kept constant when going to the relativistic domain.
By group theoretical means they obtained the wavefunctions for such a
`relativistic' oscillator, which up to phase factors and a `weight function',
contain a polynomial part dubbed by ABNS the `relativistic Hermite polynomials',
$F_n^N(\xi)$, of degree $n$,
parameter $N=mc^2/\hbar\omega$ (considered as discrete) and variable
$\xi _N=(\omega/c)\sqrt{N}x$. In this paper, we shall call them the
ABNS functions, because Nagel \cite{3}
showed that they actually are a product of a factorial, a simple polynomial,
and the Gegenbauer \p of the same degree and parameter but of a different
argument.

In the ABNS paper, recurrence relations with respect to the polynomial degree
$n$ are written down.
One might think of similar relations for $N$, which, however, appears both
as a polynomial parameter and as a discrete factor in the argument of the ABNS
functions. We have here a case of both shift relations with respect to a
discrete variable from the point of view of the ABNS functions and recurrence
relations from the point of view of the polynomial parameter.
We recall that shifts with respect to a continuous variable are a common
procedure for relativistic oscillators of finite difference type \cite{russ}.
In the following,
we shall show that the recurrence-shift relations for the ABNS functions
can be obtained almost
trivially from the corresponding recurrence relations of the Gegenbauer
polynomials. Moreover, we shall give some hints on the
physical relevance of these recurrence-shift relations,
which from their definition appear to be related to pair
creation processes. A very short presentation of our results has been
published in the Proceedings of Wigner lV Symposium \cite{wig}, and here one
may find a more detailed study.
We mention that Zarzo et al. \cite{zm} studied the asymptotic
distribution of the zeros of ABNS functions, and other
algebraic and spectral properties.
The ABNS functions satisfy the following second-order linear differential
equation
$$
(1+\xi _N ^2/N)y^{''}-(2/N)(N+n-1)\xi _Ny^{'} + (n/N)(2N+n-1)y =0~,
\eqno(1)  $$
where the derivatives are with respect to the $\xi _N$ variable.
The limit $N\rightarrow \infty$ is the `nonrelativistic one',
$c\rightarrow \infty$, in which
the ABNS functions go into the usual Hermite \pp.

The base for getting our results is a simple yet sufficient general
factorization method that can be inferred
from the factorization techniques in the book of Miller \cite{mil}.
A particular case has been used by Pi\~na \cite{pi} to work out many
examples.

\section*{2.~The \f scheme}

Given the following family of second order linear differential equations
$$
P(\xi)\frac{d^2 y_s(\xi)}{d\xi ^2}+Q_s(\xi)\frac{dy_s(\xi)}{d\xi} + R_s(\xi)
y_s(\xi)=0~,
\eqno(2)
$$
let us suppose that the solutions $y_s$ to (2) satisfy the following
recurrence (ladder) relations,
$$
A_s^{\pm}y_s=r_s^{\pm}y_{s\pm1}.
\eqno(3)
$$
where $A_s^{\pm}$ are first order differential operators of the type
$$
A_s^+=f_s^+\frac{d}{d\xi}+g_s^+
\eqno(4a)
$$
$$
A_s^-= f_s^-\frac{d}{d\xi}+g_s^-~.
\eqno(4b)
$$
In Eqs.~(4), $f_s^{\pm}$ and $g_s^{\pm}$ might be functions of $\xi$,
whereas in
Eqs.~(3) $r_s^{\pm}$ depend only on the parameter $s$. By means of
the ladder operators we can write down two types of second-order differential
equations
$$
A_{s+1}^-A_{s}^+y_s=r_s^+r_{s+1}^-y_s
\eqno(5a)
$$
and
$$
A_s^+A_{s+1}^-y_{s+1}=r_s^+r_{s+1}^-y_{s+1},
\eqno(5b)
$$
involving the same constant $k_s=r_s^+r_{s+1}^-$.

By substituting Eqs.~(4a,b) in Eqs.~(5a,b), respectively,
one should obtain
the second-order linear differential equation~(2),
for the indices $s$ and $s+1$,
respectively. This allows to get the following five equations as sufficient for
Eqs.~(2) and (5) to have the same solutions
$$
P=f_s^+f_{s+1}^-
\eqno(6a)
$$
$$
Q_{s+1}=f_s^+\frac{df_{s+1}^-}{d\xi}+f_s^+g_{s+1}^-+f_{s+1}^-g_s^+
\eqno(6b)
$$
$$
Q_{s}=f_{s+1}^-\frac{df_{s}^+}{d\xi}+f_{s+1}^-g_{s}^++f_{s}^+g_{s+1}^-
\eqno(6c)
$$
$$
R_{s+1}+k_s=f_{s}^+\frac{dg_{s+1}^{-}}{d\xi}+g_s^+g_{s+1}^{-}
\eqno(6d)
$$
$$
R_{s}+k_s=f_{s+1}^-\frac{dg_{s}^{+}}{d\xi}+g_s^+g_{s+1}^{-}~.
\eqno(6e)
$$
Taking the derivative of $P$ with respect to $\xi$ and subtracting
Eq.~(6c)
from (6b) one will obtain
$$
\frac{1}{2}\left(\frac{dP}{d\xi}+Q_{s+1}-Q_s\right)= f_s^+
\frac{df_{s+1}^-}{d\xi}=P\frac{d}{d\xi}\ln f_{s+1}^-~.
\eqno(7)
$$
From Eq.~(7) one gets easily
$$
f_{s+1}^-=\sqrt{P}\exp\Bigg[\frac{1}{2}\int\frac{Q_{s+1}-Q_s}{P}d\xi \Bigg]
=\sqrt{P}E_s
\eqno(8a)
$$
and using Eq.~(6a)
$$
f_s^+=\sqrt{P}\exp\Bigg[\frac{-1}{2}\int\frac{Q_{s+1}-Q_s}{P}d\xi \Bigg]=
\sqrt{P}E_s^{-1}~.
\eqno(8b)
$$
To get $g_s^+$ and $g_s^-$ one should proceed as follows. Firstly, one can
obtain easily
$$
\frac{R_{s+1}-R_s}{\sqrt{P}}= E_s^{-1}\frac{dg_{s+1}^-}{d\xi}-E_s
\frac{dg_s^+}{d\xi}
\eqno(9)
$$
and
$$
\frac{1}{2}\left(Q_{s+1}+Q_s-\frac{dP}{d\xi}\right)\equiv \sqrt{P}W_s=
f_s^+g_{s+1}^- + f_{s+1}^-g_s^+~.
\eqno(10)
$$
It follows
$$
W_s=E_s^{-1}g_{s+1}^-+E_sg_s^+~.
\eqno(11)
$$
After some algebra one gets
$$
\frac{R_{s+1}-R_s}{\sqrt{P}}+\frac{dW_s}{d\xi}= \frac{d}{d\xi}
\left(2E_s^{-1}g_{s+1}^-\right)
+W_s\frac{d\ln{ E_{s}}}{d\xi}~.
\eqno(12)
$$
Thus
$$
g_{s+1}^-=\frac{E_s}{2}\Bigg[W_s+\int\left(\frac{R_{s+1}-R_s}{\sqrt{P}}
-W_s\frac{Q_{s+1}-Q_s}{2P}\right)d\xi \Bigg]~.
\eqno(13)
$$
The other coefficient $g_s^+$ is obtained from
$g_s^+=\frac{W_s}{E_s}-\frac{g_{s+1}^-}{E_s^2}$ giving
$$
g_s^+=\frac{1}{2E_s}\Bigg[W_s-\int\left(\frac{R_{s+1}-R_s}{\sqrt{P}}
-W_s\frac{Q_{s+1}-Q_s}{2P}\right)d\xi \Bigg]~.
\eqno(14)
$$
Finally, the $k$-constant can be found from the most convenient of
Eqs.~(6d,e). In order to find the $r$-coefficients one needs either
supplementary
information on the \p solutions of second order linear differential
equations or some tricks as one can see in the applications to
follow.

\section*{3.~Application to ABNS polynomial functions}

3.1. {\em Factorization with respect to the \p degree}

In this case we have $s=n$, $P(\xi)=1+\xi _N^2/N$,
$Q_n=-2(N+n-1)\xi _N/N$, $R_n=n(2N+n-1)/N$. The calculations are
straightforward. However we present some details. The convenient variable to
work with is $u=\xi _N/\sqrt{N}$. Other variables will change by logarithmic
terms the constants of integration, which anyway we shall not include here
as far as we are interested in the simplest solutions. The first thing to do
is to calculate the integral in the exponential for the $f$ coefficients,
which reads
$$\int \frac{Q_{n+1}-Q_n}{P}d\xi=\int \frac{-2\xi d\xi}{N+\xi ^2}=
\int \frac{-2u du}{1+u^2}=-\ln{(1+u^2)}
\eqno(15)
$$
and by substituting in Eqs.~(8a,b) one gets $f^-=1$ and
$f^+=1+u^2$, respectively, both not depending on $n$.
$W_n$ can be obtained from the
identity in Eq.~(10) leading to
$W_n=-\frac{2(N+n)}{\sqrt{N}} \frac{u}{\sqrt{1+u^2}}$.
The first term in the integrals for $g^{\pm}$ reads
\newpage
$$
\int\frac{R_{n+1}-R_n}{\sqrt{P}}d\xi=\frac{2(N+n)}{\sqrt{N}}
\ln \left(u+
\sqrt{1+u^2}\right)=
$$
$$\frac{2(N+n)}{\sqrt{N}} {\rm arcsinh} u~.
\eqno(16)
$$
The second part of the integral in Eqs.~(13, 14) is amenable to the
following simple one
$$
\frac{2(N+n)}{\sqrt{N}}\int \frac{u^2 du}{(1+u^2)^{3/2}}=
\frac{2(N+n)}{\sqrt{N}}\Bigg[\ln\left(u+\sqrt{1+u^2}\right)-
\frac{u}{\sqrt{1+u^2}}\Bigg]=$$
$$\frac{2(N+n)}{\sqrt{N}}\Bigg[{\rm arcsinh} u-
\frac{u}{\sqrt{1+u^2}}\Bigg] ~.
\eqno(17)
$$
The final result for the $g$-terms is
$$
g^{\mp}=\frac{(N+n)}{\sqrt{N}}(\sqrt{1+u^2})^{\mp 1}
\Bigg[-\frac{u}{\sqrt{1+u^2}}\pm
{\rm arcsinh} u
\mp
{\rm arcsinh} u
\pm \frac{u}{\sqrt{1+u^2}}\Bigg]~.
\eqno(18)
$$
So, in this case, $g_{n+1}^-=0$ and $g_{n}^{+}=-2(1+\frac{n}{N})\xi _N$.
Since $g^-$ is naught, from Eq.~(6d) one will find out that
$k_n=-R_{n+1}$.

The recurrence relations for ABNS functions can be written as follows
$$
A_n^+y_n\equiv
\Bigg[(1+\frac{\xi _N ^2}{N})\frac{\partial}{\partial \xi _N}-
2(1+\frac{n}{N})\xi _N\Bigg]y_n=
r_n^+y_{n+1}
\eqno(19a)
$$
and
$$
A_n^-y_n\equiv\frac{\partial}{\partial \xi _N}y_n=r_n^-y_{n-1}~.
\eqno(19b)
$$
In order to proceed further we
have to be aware of the fact that the \f scheme above
does not allow to find out the $r$-coefficients, but only the $k$-constant,
i.e., their product. As a matter of fact, the ambiguities of \f procedures
have been known since Infeld and Hull \cite{ih}.
For the present method, the $k$-constant comes out quite often in
factorized form and then one can make some selection of the
$r$-coefficients identified with the $k$-factors (though this is not a rule)
on the base of further criteria.
This situation is clearly illustrated by the ABNS functions.
Indeed, from Eq.~(6d) one gets $k_n=-R_{n+1}=-\frac{(n+1)(2N+n)}{N}$.
Because of the
three factors contained in $R$, to which one should add the minus sign,
there are sixteen $r$-pairs leading to the same $k$-constant.
However, if one asks for the first two ABNS functions to be identical to the
first two Hermite polynomials, i.e., $F_0^N=1$ and $F_1^N=2\xi _N$,
respectively, one can show easily that $r_0^+=-1$ and $r_1^-=2$.
In this way most of the $r$-pairs are discarded,
and one ends up with just two cases, i.e., ({\it i}) $r_n^+=-1$ and
$r_{n+1}^-=\frac{(n+1)(2N+n)}{N}$, ({\it ii}) $r_n^+=-(n+1)$ and
$r_{n+1}^-=(2N+n)/N$. However the latter pair is merely a rescaling of
the pure numerical coefficients entering the ABNS functions.
On the other hand,
the first pair corresponds exactly to the
recurrence relations obtained by ABNS (see Eqs.~(7)
and (8) in their paper, which are our Eqs.~(19a) and~(19b),
respectively, when the first $r$-pair is used).

\vskip 0.1cm

3.2. {\em Factorization with respect to the parameter}

To write down recurrence relations with respect to the parameter $N$ seems
impossible from the point of view of the factorization scheme
because it occurs also in the variable of the ABNS functions
and consequently one deals in fact with recurrence-shift relations
involving ABNS functions of different
variables. Therefore a direct application of our method is not possible.
We have found a way to obtain such relations by making use of the
aforementioned
result of Nagel \cite{3} who proved that ABNS functions are a product of a
factorial, a simple polynomial and a Gegenbauer \p.
Thus,
the idea is first to obtain recurrence relations for the Gegenbauer
polynomials with respect to
their parameter, which fits in our \f scheme since the Gegenbauer
variable is not a $N$-depending quantity. More exactly
we derive recurrence relations for the equation
$$
x^2(1-x^2)y''-(2N+1)x^3y'+n(2N+n)x^2y=0~.
\eqno(20)
$$
Then $P=x^2(1-x^2)$, $Q_N=-(2N+1)x^3$, $R_N=n(2N+n)x^2$. The application of
the \f scheme is straightforward leading to the following
factorizing coefficients $f^+=x$, $f^-=x(1-x^2)$, $g^+_N=2N+n$,
$g^-_{N+1}=-(2N+n+1-nx^2)$. The constant $k_N=-(2N+n+1)(2N+n)$ is in factored
form but we shall not make a direct identification of the $r$-coefficients
with the $k$-factors. Instead we take $r_N^+=2N$ and thus $r_{N+1}^-=
-(2N+n+1)(2N+n)/2N$. The recurrence relations read
$$
\Bigg[x\frac{\partial}{\partial x}+(2N+n)\Bigg] C_n^N=2NC_n^{N+1}~
\eqno(21a)
$$
and
$$
\Bigg[x(1-x^2)\frac{\partial}{\partial x}-(2N+n-1-nx^2)\Bigg]C_n^N =
-\frac{(2N+n-1)(2N+n-2)}{2N-2}C_n^{N-1}~.
\eqno(21b)
$$
The first one can also be obtained by combining Eqs.~(8.933.3) and
(8.935.2)
in Gradshteyn and Ryzhik \cite{gr}, while the latter one can be reached
from the set (8.933.2-4), (8.935.2) in the same reference.

Nagel \cite{3} proved the following relationship
$$
F_n^N(u\sqrt{N})=n!\left(\frac{1+u^2}{N}\right)^{n/2}
C_n^N(u/\sqrt{1+u^2})~.
\eqno(22)
$$
Plugging Nagel's result into equations~(21a,b) one gets the following
recurrence-shift relations for the ABNS functions
$$
\Bigg[\xi _N\left(1+\xi _N^2/N\right)\frac{\partial}{\partial \xi _N}
-\frac{n}{N}\xi _N^2+2N+n\Bigg]
F_n^N(\xi _N)=$$
$$2N\left(\sqrt{1+\frac{1}{N}}\right)^n
F_n^{N+1}\left(\sqrt{1+\frac{1}{N}}\xi _N\right)
\eqno(23a)
$$
and
$$
\Bigg[\xi _N\frac{\partial}{\partial \xi _N}-(2N+n-1)\Bigg]F_n^N(\xi _N)=$$
$$
-\frac{(2N+n-1)(2N+n-2)}{2N-2}\left(\sqrt{1-\frac{1}{N}}\right)^n
F_n^{N-1}\left(\sqrt{1-\frac{1}{N}}\xi _N
\right)~.
\eqno(23b)
$$
Have these relationships any application in physics ?
Our answer is as follows. The
wavefunctions containing the ABNS polynomial functions satisfy
the Klein-Gordon equation associated with the Anti-de Sitter (AdS) metric
(of negative curvature $R=-\omega ^2/c^2$) as stated by ABNS
$$
(\Box +m^2c^2/\hbar ^2+NR)\psi=0~.
\eqno(24)
$$
The true limiting process (or contraction in group theory terminology)
should be taken as
$c\rightarrow\infty$ and simultaneously
$R=-\omega ^2/c^2\rightarrow 0$, keeping $c\sqrt{|R|}=\omega$, leading from
the AdS relativity to the Newton one \cite{gar}. The AdS symmetry group is a
deformation of both the relativistic free particle and the harmonic oscillator.
At the quantum level, the double covering SU(1,1) of the AdS group is
represented by the discrete series of representations, which are indexed by the
dimensionless number $\eta =N+\frac{1}{2} + O(\sqrt{|R|}$ \cite{gar}.
Our claim is that the recurrence-shift relations of ABNS oscillator functions
can be important in the spatial analysis of the
pair-production effects in a wave approach in the AdS
gravitational background, since their action is of connecting
ABNS functions
with consecutive parameters related to discrete changes in the
gravitational curvature (in the Klein-Gordon equation, Eq.~(24),
$N$ is like a discrete coupling constant for the curvature)
and simultaneously discrete changes in the
polynomial variable. Indeed, one can see that $N=\frac{1}{k\lambda _C}$,
where $k$ is the usual wavevector and $\lambda _C$ is the Compton wavelength.
Thus, one can think of $N$ as the ratio of the spatial resolution of a common
wave
$\propto 1/k$ and the spatial resolution given by the Compton wavelength.
We recall the similar suggestion of Noll \cite{n} in Optics on the
usefulness of ladder operations of Zernike polynomials, whenever the gradient
of a wave front is required. Moreover, to be recalled is the idea of hadrons
as AdS microuniverses of huge curvature \cite{ads}. Pair creation
processes in such a context might be examined in the aforementioned
perspective as well.


\section*{Acknowledgments}
This work was partially supported by the CONACyT Projects 4862-E9406 and
4868-E9406 and a CONACyT Graduate Fellowship.

\end{document}